\documentclass[prd,nofootinbib,preprint]{revtex4}
\usepackage{graphicx}
\usepackage{relsize}
\usepackage{epsfig}
\usepackage{amsmath}
\usepackage{amsfonts}
\usepackage{amssymb}
\usepackage{url}
\usepackage{hyperref}
\usepackage{subfigure}
\usepackage{hhline}
\usepackage{color}
\usepackage{bm}
\usepackage{cancel}
\newcommand{\beqa}{\begin{eqnarray}}
\newcommand{\eeqa}{\end{eqnarray}}
\newcommand{\beq}{\begin{equation}}
\newcommand{\eeq}{\end{equation}}

\newcommand{\bmt}{\begin{pmatrix}}
\newcommand{\emt}{\end{pmatrix}}
\usepackage[toc,page]{appendix}
\usepackage{comment}
\usepackage{hyperref}
\usepackage{epstopdf}
\newcommand{\be}{\begin{equation}}
\newcommand{\ee}{\end{equation}}
\newcommand{\bea}{\begin{eqnarray}}
\newcommand{\eea}{\end{eqnarray}}

\begin{document}
\title{Constraining CPT violation with Hyper-Kamiokande and ESSnuSB}
\author{Rudra Majhi$^{1}$}
\email{rudra.majhi95@gmail.com}

\author{Dinesh Kumar Singha$^{1}$}
\email{dinesh.sin.187@gmail.com}

\author{K. N. Deepthi$^{2}$}
\email{nagadeepthi.kuchibhatla@mahindrauniversity.edu.in}

\author{Rukmani Mohanta$^{1}$}
\email{rmsp@uohyd.ac.in}

\affiliation{$^1$\,School of Physics, University of Hyderabad,
              Hyderabad - 500046, India \\                   
 $^2$\,School of Physics, Mahindra University, Hyderabad - 500043, India }

\begin{abstract}
CPT invariance is one of  the most fundamental symmetries in nature and it plays a  major role in the formulation of Quantum Field Theory. Although no definitive signal of CPT violation has been observed so far, there are many reasons to   carefully investigate  various low-energy phenomena that can provide better probes to test CPT symmetry. In this context, neutrino experiments are expected to provide more stringent bounds on CPT invariance violation when compared to the existing bounds from the Kaon system.  
In this work, we investigate the sensitivity of the upcoming long-baseline experiments: Hyper Kamiokande (T2HK, T2HKK),  ESSnuSB and DUNE to constrain the CPT violating parameters  $\Delta(\delta_{CP})$, $\Delta(m^2_{31})$ and $\Delta(\sin^2 \theta_{23})$, which characterize the difference between neutrino and antineutrino oscillation parameters. Further, we analyse neutrino and antineutrino data independently and constrain the oscillation parameters governing them by considering the combination of these experiments (DUNE+T2HKK and DUNE+ESSnuSB). In addition, assuming CPT symmetry is violated in nature, we study the individual ability of the aforementioned experiments to establish CPT violation. We found that the  experiments Hyper-K (T2HK, T2HKK) and ESSnuSB, along with DUNE, will be able to establish CPT violation in their proposed run-times.
\end{abstract}
\maketitle
 
\section{Introduction}

Understanding the physics beyond Standard Model (BSM) is one of the prime objectives of present-day particle physics research. With the non-observation of any new heavy BSM particle through direct detection at LHC, the focus has been shifted to other frontiers, e.g., Intensity and Cosmic. In the  Intensity frontier, neutrinos provide a promising avenue for revealing  new physics. The compelling  evidence of neutrino oscillations from various experiments  already indicates that the minimal Standard Model (SM)  of particle physics is not exhaustive and requires modification. In general, SM is considered as a low-energy effective theory originating from the unified theory of Quantum Gravity at the Planck scale. Hence,  understanding the true nature of the Planck scale physics through experimental signatures  is of great importance, albeit extremely challenging to identify.  It is expected   that the  long-baseline experiments will provide the ideal  platform to look for tiny violations of Lorentz invariance or CPT symmetry that may exist as the low-energy remnants of Planck scale physics.

It is well-established that local relativistic quantum field theories, including the Standard Model, are  invariant under Lorentz and CPT transformations. CPT theorem \cite{CPT-theorem} states that ``\textit{Any quantum theory formulated on flat space time is symmetric under the combined action of CPT transformations, provided the theory respects (i) Locality (ii) Unitarity and (iii) Lorentz invariance}". One of the phenomenological consequences of CPT symmetry is that  particles and antiparticles will have the same masses and lifetimes. If  any discrepancy is found either in their masses or lifetimes, it would be a clear sign of CPT violation.
The results from numerous experiments are  consistent with the predictions of this symmetry. Although no conclusive evidence of CPT violation has been observed so far, there are many reasons to perform a careful investigation of  possible mechanisms
and descriptions of Lorentz and CPT violations. One of the ambitious motivations
is that the  Lorentz and CPT violations might arise from a fundamental theory at the Planck scale, but nonetheless may leave their footprints in some low-energy observables which can be detected in the current or upcoming experiments of exceptional sensitivity. 

Studies related to CPT violation are not new, see e.g., Refs. \cite{CPT10,Baren,CPT11,CPT12,CPT13,CPT14,CPT15,CPT16,CPT17,CPT18,CPT19, CPT19a,CPT20,CPT21,CPT22, Baren-1, LIV, CPTV-Daljeet}. There are several theories by which Quantum Gravity induced CPT violation can occur. Especially neutrinos, in addition to neutral kaons \cite{Schwingenheuer:1995uf}, make potential candidates to provide good insight into CPT violation, if it exists. For instance, some interesting aspects of Quantum gravity decoherence (non-local) in neutrinos introduce CPT violation and account for the smallness of neutrino mass \cite{Capolupo}.

There exist experimental limits on CPT violating parameters from kaon and the lepton sectors. However, the current neutrino  oscillation data provides the most stringent constraints on various oscillation parameters \cite{Baren-1}: 
 \begin{eqnarray}
 && |\Delta m_{21}^2-\Delta \overline{m}_{21}^2| < 4.7\times 10^{-5} ~\text{eV}^2,
  \nonumber \\
  && |\Delta m_{31}^2-\Delta \overline{m}_{31}^2| < 2.5\times 10^{-4} ~\text{eV}^2,
 \nonumber \\
 & & |\sin^2\theta_{12}-\sin^2\overline{\theta}_{12}| < 0.14,
  \\
  && |\sin^2\theta_{13}-\sin^2\overline{\theta}_{13}| < 0.029,
  \nonumber \\
&  & |\sin^2\theta_{23}-\sin^2\overline{\theta}_{23}| < 0.19\nonumber .
 \label{eq:new-bounds}
 \end{eqnarray}
Further, in Ref.~\cite{Baren} it has been shown that DUNE will test the CPT violation in atmospheric mass difference to an unprecedented level and provide the most stringent limit as $\big |\Delta m_{31}^2 -\Delta\overline{m}_{31}^2 \big | < 8.1 \times 10^{-5}~{\rm eV}^2$ at $3 \sigma$ C.L.  

 Without delving into the specifications of any model, in this work we would like to test the predictions of CPT conservation in the light of future neutrino oscillation experiments Hyper-Kamiokande (T2HK and T2HKK), the European Spallation Source $\nu$-Beam (ESSnuSB) project and Deep Underground Neutrino Experiment (DUNE). Since neutrino oscillation experiments are only sensitive to mass-squared differences and mixing angles, one can test the CPT symmetry by measuring the differences in the oscillation parameters of neutrinos and antineutrinos. If the fundamental CPT invariance is not assumed, neutrinos and antineutrinos need to be parameterized by  different $3 \times 3$ unitary mixing matrices. In the case of neutrinos, the flavor eigenstates $|\nu_\alpha \rangle $ and the mass eigenstates $|\nu_i\rangle$ are related by a $3 \times 3$ unitary leptonic mixing matrix  \cite{Bilenky:1987ty},
\begin{equation}
|\nu_\alpha \rangle = \sum^3_{i=1} U_{\alpha i}(\theta_{12}, \theta_{13}, \theta_{23}, \delta_{CP}) |\nu_i\rangle\;.
\label{eq:nu}
\end{equation}
Analogously, the corresponding states for the  antineutrinos are related as
\begin{equation}
|\overline{\nu}_\alpha \rangle = \sum^3_{i=1} U^*_{\alpha i}(\overline{\theta}_{12}, \overline{\theta}_{13}, \overline{\theta}_{23}, \overline{\delta}_{CP}) |\overline{\nu}_i\rangle\;.
\label{eq:nubar}
\end{equation}
Denoting the neutrino and antineutrino masses  by $m_i$ and $\overline{m}_i$ ($i = 1, 2, 3$),  the  mass-squared differences of neutrinos are represented as 
$\Delta {m}^2_{ij} \equiv {m}^2_i -{m}^2_j$ and that of antineutrinos  as
$\Delta \overline{m}^2_{ij} \equiv \overline{m}^2_i - \overline{m}^2_j$.  Consequently the oscillation probabilities of neutrinos and antineutrinos are functions of the oscillation parameters ($\theta_{12}, \theta_{13}, \theta_{23}, \Delta {m}^2_{21}, \Delta {m}^2_{31},\delta_{CP}$) and ($\overline{\theta}_{12}, \overline{\theta}_{13}, \overline{\theta}_{23}, \Delta \overline{m}^2_{21}, \Delta \overline{m}^2_{31}, \overline{\delta}_{CP}$) respectively. In principle, neutrino oscillation experiments will be able to place bounds on the predictions of CPT symmetry violation. In this work, we investigate the ability of the future long-baseline experiments: T2HK, T2HKK, ESSnuSB and DUNE to constrain the CPT violating parameters, such as  $\left|\delta_{CP} - \overline{\delta}_{CP}\right|$, $\left|\Delta m^2_{31} - \Delta \overline{m}^2_{31}\right|$ and $\left|\sin^2 \theta^{}_{23} - \sin^2 \overline{\theta}^{}_{23}\right|$. We further,  analyse neutrino and antineutrino data independently and constrain the oscillation parameters  by considering the combination of the experiments DUNE+T2HKK and DUNE+ESSnuSB. In addition, assuming CPT symmetry is violated in nature, we study the individual ability of T2HK, T2HKK, DUNE and ESSnuSB experiments to establish CPT violation. 

%

The outline of the paper is as follows. 
In section \ref{sec:simulatn}, we give a brief overview of the experimental and simulation details of T2HK, T2HKK, ESSnuSB and DUNE. In section \ref{sec:results}, we determine the bounds placed by these experiments on the parameters $\Delta(\delta_{CP})$, $\Delta(m^2_{31})$ and $\Delta(\sin^2 \theta_{23})$ by assuming CPT symmetry exists in nature. Further, in section \ref{subsec:combined} we analyse the combined data of DUNE+T2HKK, DUNE+ESSnuSB and how well they can measure neutrino and antineutrino oscillation parameters independently. Additionally, in section \ref{sec4} we assume that CPT symmetry is violated in nature and estimate the sensitivity of T2HK, T2HKK, ESSnuSB and DUNE to establish CPT invariance violation individually. Finally, our results are summarized in section \ref{sec:summary}.
 


\section{Experimental and Simulation Details}
\label{sec:simulatn}

In this section, we  discuss the detailed experimental features of the long-baseline experiments 
T2HK, T2HKK, ESSnuSB and DUNE.\\

\textit{T2HK}: Tokai to Hyper Kamiokande (T2HK) is an upgradation proposed to the existing T2K facility in Japan. In this plan, the JPARC beam will produce a 1.3 MW powered beam and the far detector (FD) will have two identical water Cherenkov detectors of 187 kt ($2 \times 187 = 374$  kt) fiducial volume to be placed at 295 km baseline, $2.5^{\circ}$ off from the beam axis.

\textit{T2HKK}: T2HKK is an alternative choice to T2HK, where the proposed FD is placed in Korea, which is 1100 km away from the JPARC facility. One of the two tanks (187 kt) proposed in the T2HK experiment will be placed at 1100 km with an off-axis angle (OAA) of $1.5^{\circ}$ or $2^{\circ}$ or $2.5^{\circ}$. Basing on the study in \cite{CPT23}, we consider the OAA $1.5^{\circ}$ as it provides maximum sensitivity to the oscillation parameters.
We consider the proposed run time ratio of ($1\nu:3\bar{\nu}$) years corresponding to a total exposure of $27 \times 10^{21}$ protons on target (POT). The detector systematics are taken as per the \cite{CPT23, CPT24}.

\textit{ESSnuSB}: The major objective of the European Spallation Source $\nu$-Beam (ESSnuSB) project \cite{ESS} is to measure the leptonic CP violation. A neutrino beam with a peak energy of 0.25 GeV is produced at the ESS facility in Lund, Sweden. This beam is made to travel 540 km to encounter a water Cherenkov detector of 500 kt to be placed at a mine in Garpenberg. The proposed runtime is ($2\nu+8\bar{\nu}$) years with a total  POT of $27 \times 10^{22}$ corresponding to a 5 MW proton beam.

\textit{DUNE}: The Deep Underground Neutrino Experiment (DUNE) \cite{DUNE} comprises of a broad band neutrino beam of 0.5-8 GeV energy, a near detector (ND) at Fermilab and a liquid argon time projection chamber (LArTPC) of fiducial volume 40 kt located at 1300 km in South Dakota. We have considered ($5\nu+5\bar{\nu}$) year run-time, beam power of 1.2 MW corresponding to $10 \times 10^{21}$ protons on target. 

We have performed the numerical analysis using the GLoBES package \cite{CPT25, CPT26}.  The experimental specifications along with the signal and background normalisation errors, are listed in Table \ref{exp-details}. 

The statistical $\chi^2$ is obtained using 
\begin{eqnarray}
 \chi^2_{{\rm stat}} = 2 \sum_i \Big\lbrace N_i^{{\rm test}}-N_i^{{\rm true}}+N_i^{{\rm true}} \ln \frac{N_i^{{\rm true}}}{N_i^{{\rm test}}} \Big\rbrace, 
 \label{mh-eq}
\end{eqnarray}
where $N_i^{{\rm test}}$ corresponds to the number of events predicted by the model while $N_i^{{\rm true}}$ denotes the total number of simulated events (signal and background) in $i^{th}$ bin. The systematic uncertainties are incorporated into the simulation using the Pull method. The pull variables being the signal and background normalisation uncertainties of $\nu_{e}$, $\bar{\nu}_{e}$ appearance and $\nu_{\mu}$, $\bar{\nu}_{\mu}$ disappearance channels. The values of the normalisation errors on signals and backgrounds (bkg) corresponding to different channels of the experiments are listed in Table \ref{exp-details}. Here, $\chi^2_{{\rm pull}}$ accounts for these errors and acts as a penalty term to the total $\chi^2$  ($\chi^2_{{\rm tot}}$).

\begin{table}[h!]
\begin{center}
\begin{tabular}{|c|c|c|c|c|}
\hline
Experiment   & T2HK  & T2HKK & ESSnuSB & DUNE \\       
\hline
Baseline   & 295 km  & 295 km; 1100 km  & 540 km & 1300 km   \\
\hline
Fiducial Volume  & 374 kt  & 187 kt (@ 295 km) & & \\
 & & + 187 kt (@ 1100km) & 500 kt & 40 kt \\
\hline
Normalisation uncertainty  &  &  &  & \\
$\nu_e$ signal (bkg)        & 3.2\% (5\%)  & 3.8\% (5\%)   & 3.2\% (5\%)  &  2\% (5\%)\\
\hline
$\bar{\nu}_{e}$ signal (bkg)   & 3.9\% (5\%)  & 4.1\% (5\%)    & 3.9\% (5\%) & 2\% (5\%)\\
\hline
$\nu_\mu$   signal (bkg)    & 3.6\% (5\%)  & 3.8\% (5\%)    & 3.6\% (5\%)  & 5\% (5\%)\\
\hline
$\bar{\nu}_{\mu}$  signal (bkg)  & 3.6\% (5\%)  & 3.8\% (5\%)   & 3.6\% (5\%) & 5\% (5\%) \\
\hline
\hline
\end{tabular}
\end{center}
\caption{{\footnotesize The experimental specifications and systematic uncertainties of T2HK, T2HKK, ESSnuSB and DUNE.}}
\label{exp-details}
\end{table}

\begin{table} 
\begin{tabular}{|c|c|c|} \hline
Parameters            & True values               & Test value Range  \\ \hline
$\sin^2 \theta_{12}$  & 0.304 & NA      \\ 
$\sin^2 \theta_{13}$ & 0.02221                     & 0.02034 $\rightarrow$ 0.02430 \\ 
$\sin^2 \theta_{23} $ & 0.57                  & $0.4\rightarrow 0.62$\\ 
$\delta_{CP} $       & $ 195^\circ$                  & $0^\circ \rightarrow 360^\circ $\\ 
$\Delta m^2_{12}$    & $7.42 \times 10^{-5}~{\rm eV}^2 $ & 
NA \\ 

$\Delta m^2_{31}$    &~ $ 2.514 \times 10 ^{-3}~{\rm  eV}^2$ (NH)~~&~ $(2.431 \rightarrow 2.598) \times 10^{-3}~ {\rm eV}^2$~ \\
 \hline
\end{tabular}
\caption{The values of oscillation parameters that we considered in our analysis are taken from Ref. \cite{value}.}
\label{table-1}
\end{table}  

\section{Sensitivity of the experiments T2HK, T2HKK, ESS{\lowercase{nu}}SB to CPT violation}
\label{sec:results}

In this section, we calculate the CPT violation sensitivity of the long-baseline experiments T2HK, T2HKK, ESSnuSB and DUNE. 
At first, we assume that there is no intrinsic CPT violation in nature i.e., both neutrino and antineutrino parameters are equal. That is for a given oscillation parameter, we simulate the data for each of these experiments with $\Delta x= |x-\bar{x}|=0$, where $x$ ($\bar{x}$) is the oscillation parameter for neutrinos (antineutrinos). Then we evaluate the sensitivity of each of the experiments to non-zero $\Delta x$. The true values for oscillation parameters  used are given in Table \ref{table-1}. In each case, we choose three values for $\theta_{23}$:  lower octant ($\sin^2 \theta_{23}=0.43$), maximal ($\sin^2 \theta_{23}=0.5$) and higher octant ($\sin^2 \theta_{23}=0.57$) to study the correlation between the CPT  sensitivity and the octant of $\theta_{23}$. In the test values, we marginalize over all the oscillation parameters for both neutrinos and antineutrinos except $x$, $\bar{x}$ and the solar parameters (since T2HK, T2HKK, ESSnuSB and DUNE have no sensitivity to these parameters). After marginalization, we calculate the $\chi^2$ value for the CPT violating observable through the relation
\begin{equation}
\chi^2 (\Delta x)= \chi ^2(|x-\bar{x}|)= \chi^2(x) + \chi^2(\bar{x})\;,
\end{equation} 
and the minimum $\chi^2(\Delta x)$ has been calculated over all possible combinations of $|x-\bar{x}|$. 

\begin{figure}
\includegraphics[scale=0.31]{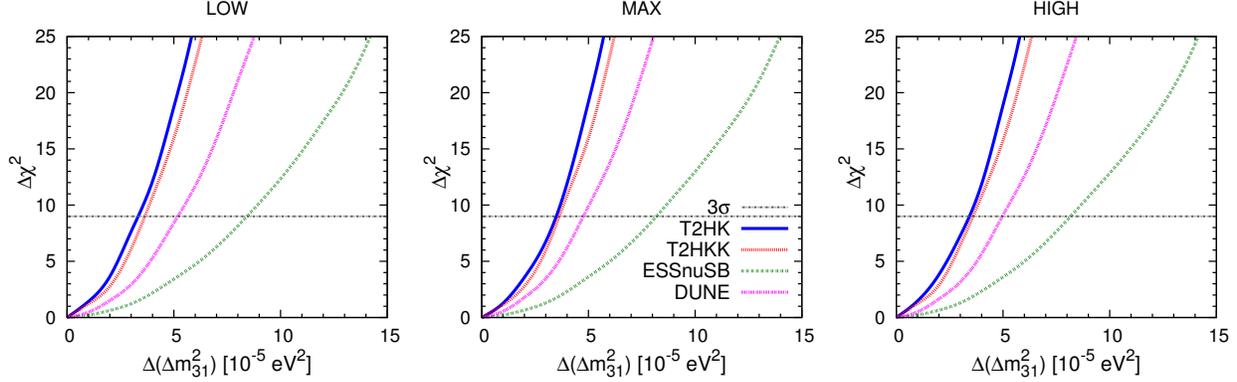}
\caption{The sensitivity of the experiments T2HK (blue curve), T2HKK (red curve), ESSnuSB (green curve) and DUNE (magenta curve) to $\Delta (\Delta m^2_{31})$.}
\label{fig-CPTV-Dm31}
\end{figure}

\begin{figure}
\includegraphics[scale=0.31]{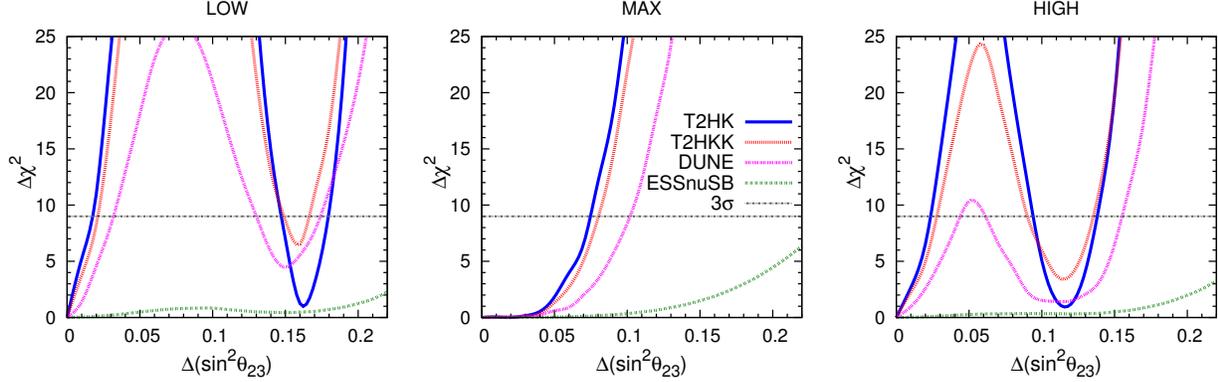}
\caption{The sensitivity of the experiments T2HK (blue curve), T2HKK (red curve), ESSnuSB (green curve) and DUNE (magenta curve) to $\Delta (\sin^2\theta_{23})$.}
\label{fig-CPTV-Dth23}
\end{figure}

\begin{figure}
\includegraphics[scale=0.31]{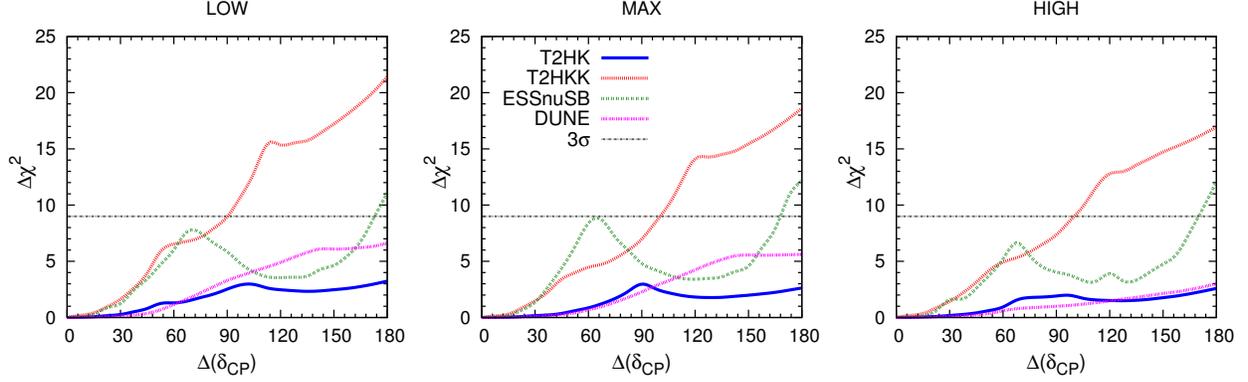}
\caption{The sensitivity of the experiments T2HK (blue curve), T2HKK (red curve), ESSnuSB (green curve) and DUNE (magenta curve) to $\Delta(\delta_{CP})$.}
\label{fig-CPTV-Ddcp}
\end{figure}

In Figs.~\ref{fig-CPTV-Dm31}, \ref{fig-CPTV-Dth23} and \ref{fig-CPTV-Ddcp}, we show the CPT violation sensitivity of these experiments to oscillation parameters $\Delta(\Delta m^2_{31}) \equiv  \Delta m^2_{31} -\Delta \overline{m}_{31}^2$, $\Delta(\sin^2 \theta_{23}) \equiv \sin^2 \theta_{23}-\sin^2\overline{\theta}_{23}$ and $\Delta(\delta_{CP}) \equiv \delta_{CP}-\bar \delta_{CP}$ respectively. The results in left, middle and right columns of these figures are obtained by assuming the octant of $\theta_{23}$ as low, maximal and high respectively. The coloured curves blue, red, green and magenta show the sensitivities of T2HK, T2HKK, ESSnuSB and DUNE respectively. The black dash-dot line represents 3$\sigma$ confidence limit.
The sensitivity for $\Delta(\sin^2\theta_{13})$ is not very significant, for which we have not shown the corresponding result here. 

From Fig.~\ref{fig-CPTV-Dm31}, one can estimate the best bound on the parameter $\Delta(\Delta m^2_{31})$ at 3$\sigma$ C.L. by T2HK experiment (blue curve) as $\Delta(\Delta m^2_{31}) < 3.32 \times 10^{-5} ~{\rm eV}^2$. It can be seen from all the columns of Fig.~\ref{fig-CPTV-Dm31} that T2HK provides a better bound compared to T2HKK, ESSnuSB and DUNE, for all the three values of $\theta_{23}$ considered. The alternative choice of the Hyper-Kamiokande experiment, i.e., T2HKK  (red curve) provides better bound on $\Delta(\Delta m^2_{31}) < 3.62 \times 10^{-5}$ ${\rm eV}^2$ at 3$\sigma$ C.L. than the bound from DUNE (magenta curve) experiment obtained in Ref.\cite{Baren} and ESSnuSB. The list of the bounds at 3$\sigma$ C.L. on $\Delta(\Delta m^2_{31})$ for $\theta_{23}<45^\circ $, $\theta_{23}=45^\circ $ and $\theta_{23}>45^\circ$ are given in the first row of Table-\ref{bounds-table}. Here it is important to note that these experiments will highly improve upon the existing bounds posed by neutral kaon system.

It can be noted from the three plots of Fig. \ref{fig-CPTV-Dth23} that different sensitivities to $\Delta(\sin^2 \theta_{23})$ are obtained for different true values of $\theta_{23}$. Firstly, when the true value of $\theta_{23}$ is in higher and lower octants, degenerate solutions are obtained for $\Delta (\sin^2\theta_{23})$ at 3$\sigma$ C.L. in the complementary octant for all the four experiments. However, for lower octant of $\theta_{23}$, this degeneracy doesn't exist in the case of T2HKK (red) and DUNE (magenta) at 2$\sigma$ C.L.. For true maximal $\theta_{23}$, the sensitivity of the experiments increase with the increasing values of $\Delta(\sin^2 \theta_{23})$. Furthermore, ESSnuSB provides comparatively  very low CPT violation sensitivity to $\Delta(\sin^2 \theta_{23})$ for all values of true $\theta_{23}$. The list of the bounds at 3$\sigma$ C.L. on $\Delta(\sin^2 \theta_{23})$ are given in second row of Table-\ref{bounds-table}.

From Fig.~\ref{fig-CPTV-Ddcp}, the best ever bounds on $\Delta(\delta_{CP})$ can be extracted from T2HKK (red curve) for CPT violation which is $\Delta(\delta_{CP})<100^0$ at $3\sigma$ confidence level. The next best bound on $\Delta(\delta_{CP})$  is obtained by ESSnuSB experiments. This is because both T2HKK and ESSnuSB experiments are planned at the second oscillation maxima to meet their primary goal of measuring the CP phase $\delta_{CP}$. The list of the bounds at 3$\sigma$ C.L. on $\Delta(\delta_{CP})$ are given in the third row of Table-\ref{bounds-table}.
\begin{table} 
\begin{tabular}{|c|c|c|c|c|} \hline
~Parameters~            & ~T2HK ~& ~T2HKK ~ & ~ESSnuSB~ &~ DUNE~ \\ \hline
$\Delta(\Delta m^2_{31})~[10^{-5}~{\rm eV}^2]$     & ~$[3.32,3.45,3.4]$ ~&  ~$[3.62,3.62,3.59]$ ~&  ~$[8.43,8.18,8.14]$ ~&   ~$[5.2,4.77,4.96]$ ~   \\
 \hline
$\Delta(\sin^2 \theta_{23})$  &   [0.180,~0.075,~0.139] & [0.167,~0.08,~0.135]&$-$ &  [0.173,~0.102,~0.155]  \\ 
\hline
$\Delta(\delta_{CP})$  &$-$ & $[90,100,100]^\circ$&$[173,168,170]^\circ$ &   $-$   \\ \hline
\end{tabular}
\caption{Bounds on the parameters at $3\sigma$ C.L. from  T2HK, T2HKK, ESSnuSB and DUNE experiments. The set of three values in the brackets correspond to  the results for $\theta_{23}$ value as $\theta_{23}<45^\circ $, $\theta_{23}=45^\circ $ and $\theta_{23}>45^\circ $. }
\label{bounds-table}
\end{table}   

\subsection{Constraining CPT violation with combination of DUNE+T2HKK and DUNE+ESSnuSB}
\label{subsec:combined}
\begin{figure}
\includegraphics[scale=0.6]{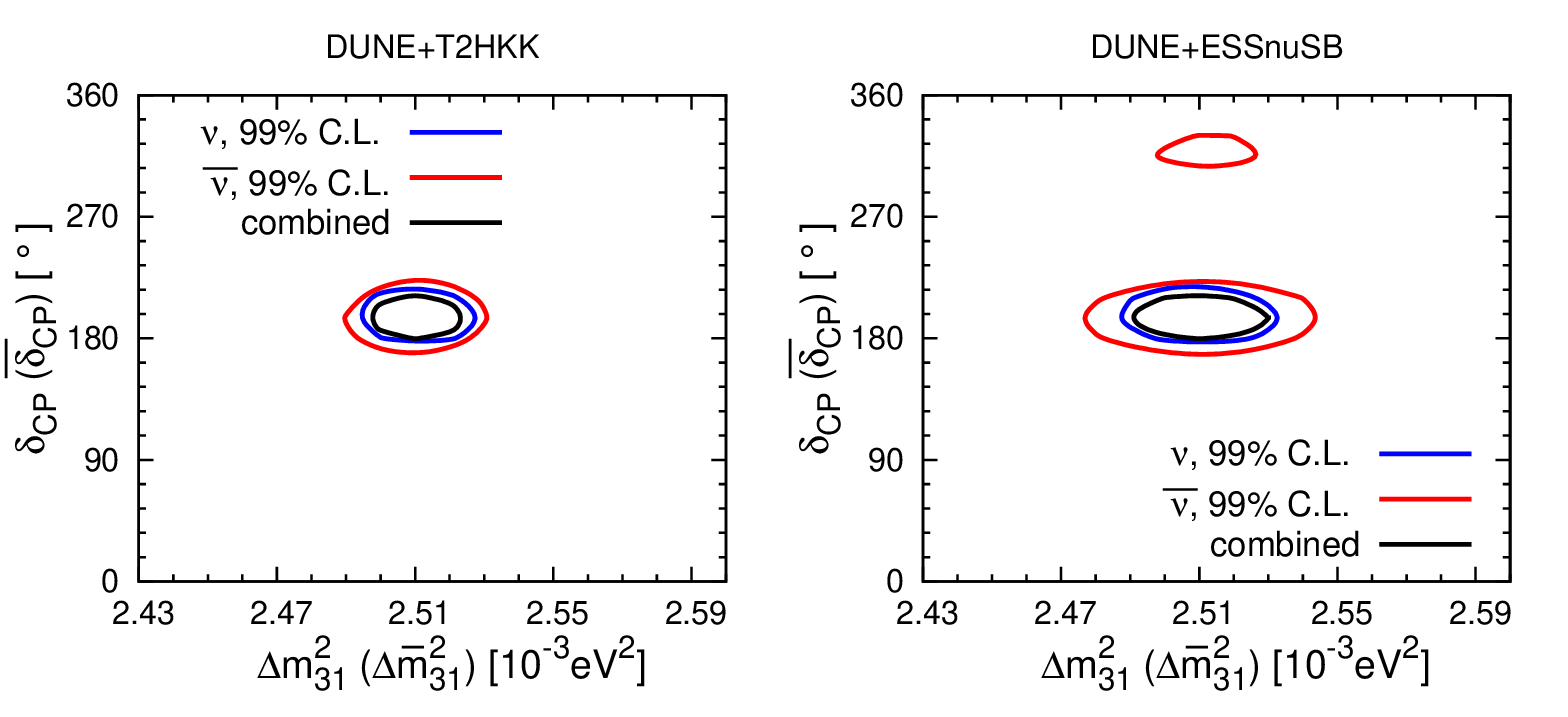}
\includegraphics[scale=0.6]{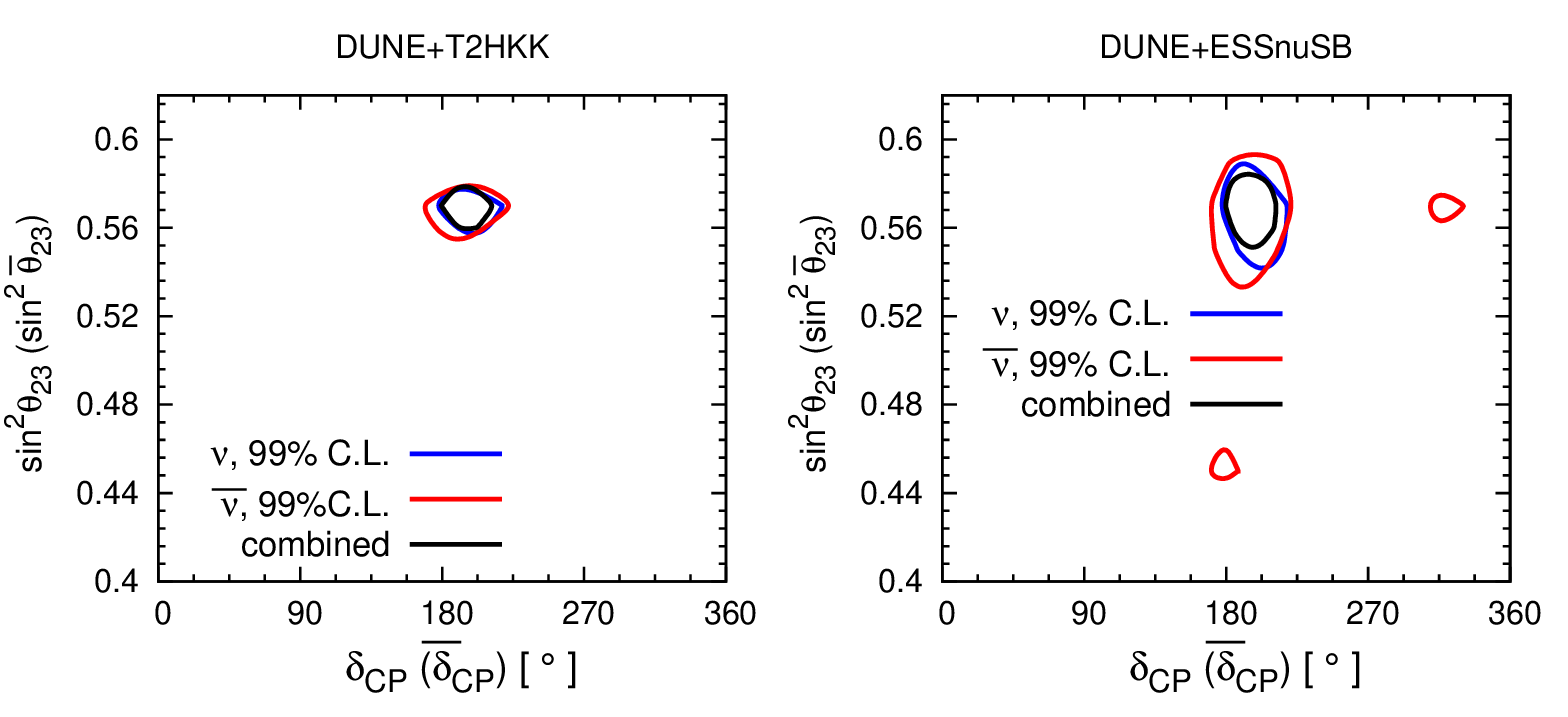}
\includegraphics[scale=0.6]{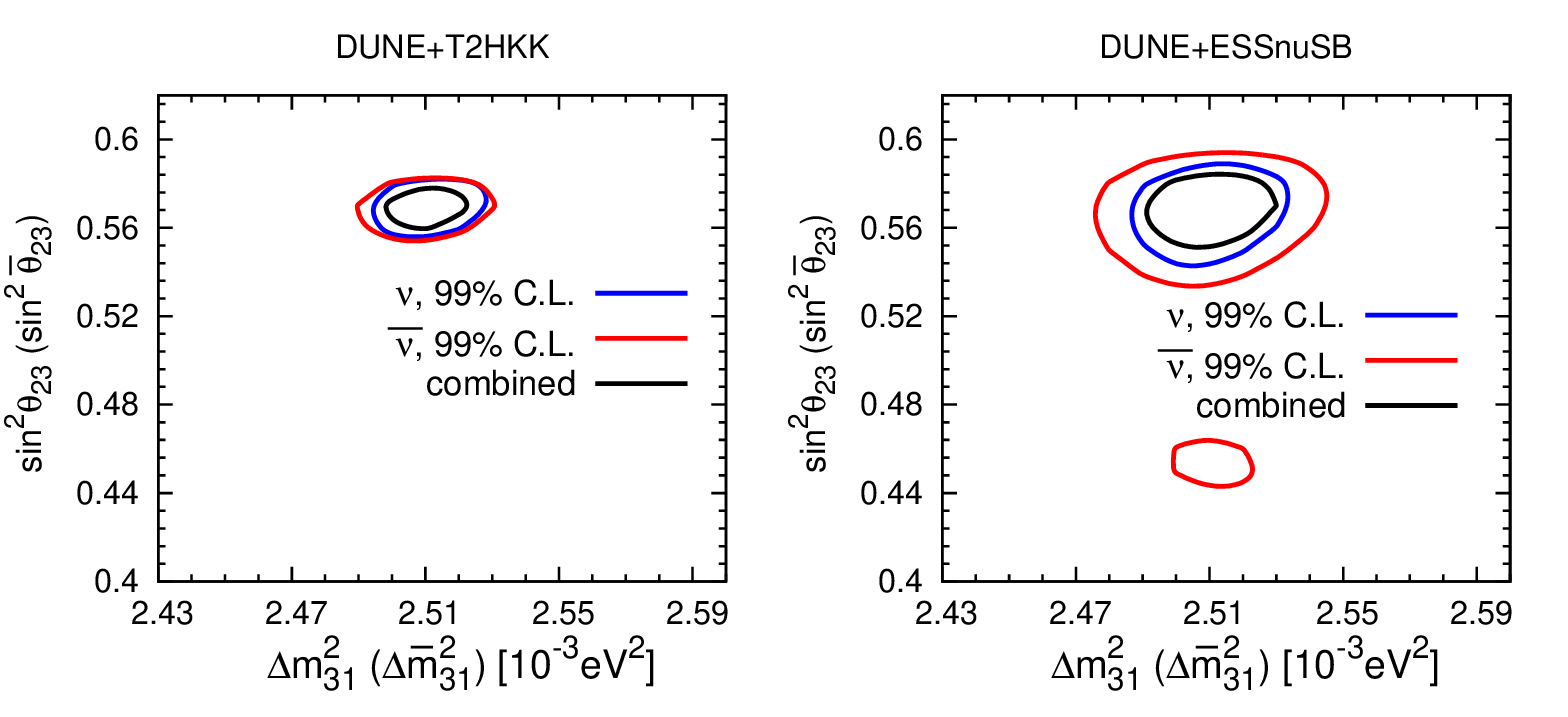}
\caption{Allowed parameter space between different neutrino and antineutrino oscillation parameters at 99$\%$ C.L. for combination of DUNE and T2HKK as well as DUNE and ESSnuSB experiments. In each plot,  the blue (red) curve is for neutrino (antineutrino) parameters and the black curve is the combined result of neutrino and antineutrino parameters.}
\label{fig-CPTV-nu-anu}
\end{figure}

In this subsection, we continue to assume that CPT is a conserved symmetry in nature. We analyse the neutrino and antineutrino data independently and determine whether the corresponding oscillation parameters in both  cases are the same as predicted by CPT symmetry. The true oscillation parameters are considered  in the analysis are provided  in Table-\ref{table-1}  and for the test scenario, we take the six oscillation parameters for both neutrino ($\Delta m^2_{31}$, $\theta_{23}$ ,$\delta_{CP}$) and antineutrino ($\Delta \overline{m}_{31}^2$, $\overline{\theta}_{23}$, $\overline{\delta}_{CP}$ ) in their allowed ranges as given in Table \ref{table-1}. Marginalisation is done over the remaining four  parameters while showing the effect of the rest  two oscillation parameters. The results are shown in Fig. \ref{fig-CPTV-nu-anu}, where the axes can be visualised for both  neutrino and antineutrino parameters.
It is shown in Ref. {\cite{andre}} that, while DUNE and T2HK can resolve the octant degeneracy assuming CPT conservation, the combination of  DUNE+T2HK cannot resolve this degeneracy while treating neutrino and antineutrino parameters individually. In this subsection, we explore the same by considering the combination of DUNE+T2HKK and DUNE+ESSnuSB experiments. The blue and red contours in all the plots in Fig.~\ref{fig-CPTV-nu-anu} represent  the allowed contours with $99\%$ C.L. for neutrino and antineutrino data respectively. The left (right) panel  shows the allowed regions of the neutrino and antineutrino oscillation parameters of DUNE+T2HKK (DUNE+ESSnuSB) experiments. From all the left panel plots, it can be seen that DUNE+T2HKK resolve octant degeneracy in $\theta_{23}$, and  $\bar{\theta}_{23}$  at $99\%$ C.L. when CPT conservation is assumed in nature. Besides, when we combined the simulated data from the antineutrino beams of DUNE and ESSnuSB (red curves of right side panel) degenerate solutions to $\delta_{CP}$ and $\theta_{23}$ are obtained. However, this degeneracy disappears when we considered the neutrino beams of DUNE and ESSnuSB. Overall, from all the plots of Fig.~\ref{fig-CPTV-nu-anu}, we can  observe that neutrino oscillation data constrains the parameters better than the antineutrino data. 

\section{Discovering CPT violation}
\label{sec4}
\begin{figure}
\includegraphics[scale=0.6]{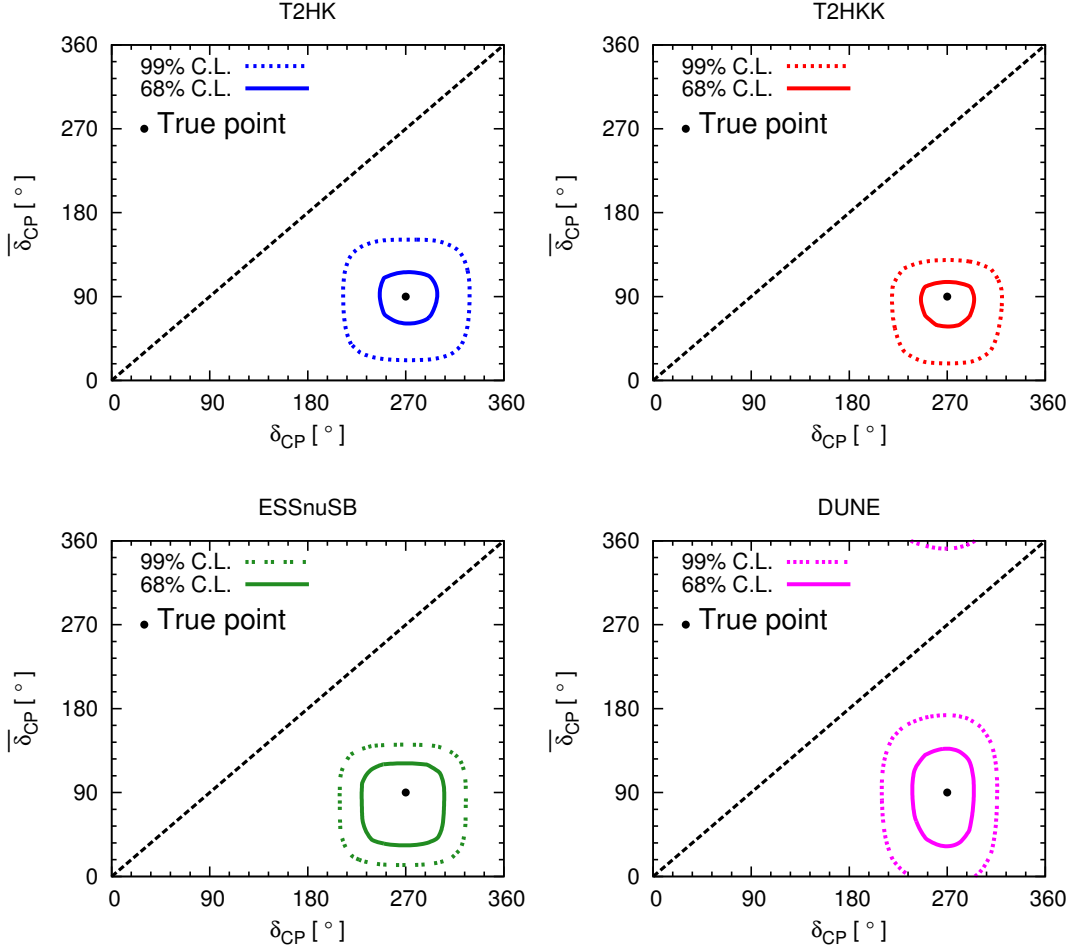}
\caption{  Allowed regions between $\delta_{CP}$ and $\overline{\delta}_{CP}$ in the  CPT violating scenario. Solid (Dotted) curve shows the parameter space at $68\%$ and $99\%$ C.L.. }
\label{fig-dcp-dcpbar}
\end{figure}

In this section, we assume that CPT is violated in nature. We generate the simulated data for the experiments T2HK, T2HKK, ESSnuSB and DUNE by assuming different neutrino and antineutrino oscillation parameters. In particular, we only consider the case where the CP-violating phases $\delta_{CP}$ and $\bar{\delta}_{CP}$ are not equal\footnote{We consider the variation in $\delta_{CP}$ and $\bar{\delta}_{CP}$ as these parameters are poorly constrained.}. We further consider $\theta_{23} = \overline{\theta}_{23}$,  $\Delta m^2_{31} = \Delta \overline{m}_{31}^2$, $\theta_{13} = \overline{\theta}_{13} $ and their true values are taken from Table-\ref{table-1}. In Fig.~\ref{fig-dcp-dcpbar}, we plot the allowed regions of $\delta_{CP}$(test) and $\bar{\delta}_{CP}$(test) as obtained from the experiments T2HK (blue), T2HKK (red), ESSnuSB (green) and DUNE (magenta). The solid and dotted contours in all the figures correspond to $68\%$ and $99\%$ C.L. and the dashed black lines correspond to CPT conserving values. Top panel of Figure~\ref{fig-dcp-dcpbar} shows that both the configurations of Hyper-Kamiokande experiment - T2HK (blue) and T2HKK (red) will be able to establish CPT violation with $99\%$ C.L. by showing that $\delta_{CP} \neq \bar{\delta}_{CP}$ in their proposed run-time. This can be inferred from the fact that there are no degenerate solutions obtained in the figure. However, T2HKK provides tighter constraints on the parameter space of $\delta_{CP}$-$\bar{\delta}_{CP}$, compared with T2HK, as seen from the red and blue contours of the top panel. This can be attributed to the higher sensitivity of T2HKK experiment to the CP violating phase $\delta_{CP}$ as its far detector (Korea) is going to be placed at second oscillation maxima ($\nu_{\mu} \rightarrow \nu_e$ channel) unlike in the case of T2HK experiment which focuses on first oscillation maxima. The bottom panel of Fig.~\ref{fig-dcp-dcpbar} shows that ESSnuSB and DUNE can establish CPT violation with $99\%$ C.L. on their own. Additionally, since ESSnuSB experiment has higher sensitivity to CP phase when compared to DUNE, it can be seen that the green contours (ESSnuSB) in the bottom left panel are tighter than the magenta contours (DUNE) of the bottom right panel. In conclusion, if nature has CPT violation, then the forthcoming experiments Hyper-K, ESSnuSB and DUNE will be able to establish CPT violation individually in their proposed run-times.
 
%
%
%

\section{Summary}
\label{sec:summary}
CPT symmetry is one of the fundamental symmetries of nature and its breaking is related to Planck scale physics. However, CPT violation has not been observed unambiguously so far, and it is an exciting challenge to search for its implications. 
It is expected that some of the CPT violating observables might be observed at low-energy scales. So far, the most stringent bound on CPT violation comes from the neutral kaon sector.  Remarkably, the current evidence for neutrino oscillations lies at levels where Planck-suppressed effects might be expected to appear.  Hence, the neutrino sector can provide a better opportunity to explore CPT violation. 

In this paper, we have studied the sensitivity reach of the upcoming long-baseline experiments, T2HK, T2HKK, ESSnuSB  and DUNE to explore the CPT violation in the neutrino sector. Our findings are summarized below:

$\bullet$ We obtained the sensitivity limits on the  CPT violating  parameters $\Delta(\Delta m^2_{31})$, $\Delta (\sin^2 \theta_{23})$ and $\Delta (\delta_{CP})$.
We found that the T2HKK and ESSnuSB experiments are quite sensitive to the CP violating phase $\delta_{CP}$, whereas T2HK, T2HKK and DUNE are sensitive to the atmospheric mixing parameters.  The most stringent limits on $\Delta(\Delta m^2_{31})$ and $\Delta (\sin^2 \theta_{23})$ come from T2HK experiment whereas T2HKK will provide the best bound on $\Delta (\delta_{CP})$. 

$\bullet$ Next, we obtained the constraint on CPT violation with the combination  DUNE+T2HKK and DUNE+ESSnuSB experiments. Assuming that nature is invariant under CPT, we analysed the neutrino and antineutrino data independently for these combinations of experiments and scrutinized whether they provide the same oscillation parameters as predicted by CPT symmetry. We found that these experiments are sensitive to CPT violation and DUNE+T2HKK can even resolve the octant degeneracy in $\theta_{23}$ and $\bar \theta_{23}$ at 99\% C.L..

$\bullet$ Finally, we have shown that if CPT violation exists in nature, the upcoming long-baseline experiments T2HK, T2HKK, ESSnuSB and DUNE  will be able to establish CPT violation individually at 99\% C.L. in their proposed run-times by demonstrating $\delta_{CP} \neq \bar \delta_{CP}$.

In conclusion, we found that the upcoming experiments T2HK, T2HKK, ESSnuSB and DUNE have great potential to establish CPT violation in neutrino oscillation and provide stringent limits on the CPT violating parameters 
$\Delta(\Delta m^2_{31})$ and $\Delta (\sin^2 \theta_{23})$.


{\bf Acknowledgements} The authors would like to thank Enrique Fernandez-Martinez for sharing the ESSnuSB GLoBES files. One of the authors (Rudra Majhi) would like to thank the DST-INSPIRE program for financial support.  DKS acknowledges  CSIR,  Govt. of India, for financial support. The work of RM is supported in part by SERB, Govt. of India through grant no. EMR/2017/001448 and Univ of Hyderabad IoE project grant no. RC1-20-012.  We gratefully acknowledge the use of CMSD HPC  facility of Univ. of Hyderabad to carry out computations in this work.

\end{document}